\documentclass[aps,prl,showpacs,twocolumn,groupedaddress,floatfix]{revtex4}
\usepackage{graphics}
\usepackage{graphicx}
\usepackage{epsfig}
\usepackage{pstricks}
\usepackage{color}

\usepackage{amsmath,amssymb}
\newcommand{\BEC}{\text{BEC}}
\newcommand{\TBECZERO}{T^{2\text{d}}_{\text{BEC}}}
\newcommand{\TBECQZERO}{T^{\text{q2d}}_{\text{BEC}}}
\newcommand{\TBECZEROTHREE}{T^{\text{3d}}_{\text{BEC}}}
\newcommand{\KT}{\text{KT}}
\newcommand{\SAT}{\text{sat}}

\newcommand{\mutilde}{\tilde{\mu}}

\newcommand{\omegatilde}{\tilde{\omega}}
\newcommand{\gtilde}{\tilde{g}}
\newcommand{\vtilde}{\tilde{v}}
\newcommand{\F}[1][2]{F_{#1}}

\newcommand{\flabel}[1]{\label{f:#1}}
\newcommand{\elabel}[1]{\label{e:#1}}
\newcommand{\eq}[1]{eq.~(\ref{e:#1})}
\newcommand{\eqq}[1]{Equation~(\ref{e:#1})}
\newcommand{\fig}[1]{fig.~\ref{f:#1}}
\newcommand{\figg}[1]{Figure~\ref{f:#1}}

\newcommand{\exph}[1]{\mathrm{e}^{#1}} 
\newcommand{\dd}[1]{\text{d}{#1\ }}   
\newcommand{\ddd}[1]{\text{d}{#1}}   
\newcommand{\expb}[1]{\exp \glb #1 \grb} 
\newcommand{\expc}[1]{\exp \glc #1 \grc} 
\newcommand{\glb}{\left(}  
\newcommand{\grb}{\right)}  
\newcommand{\glc}{\left[}  
\newcommand{\grc}{\right]}  
\newcommand{\gld}{\left\{}  
\newcommand{\grd}{\right\}}  

\begin{document}

\title{Semiclassical theory of the quasi two-dimensional trapped Bose gas}

\author{Markus Holzmann$^1$, Maguelonne Chevallier$^2$, Werner Krauth$^2$}
\affiliation{$^1$LPTMC,  Universit\'e Pierre et Marie Curie, 4 Place Jussieu,
75005 Paris, France; and LPMMC, CNRS-UJF,  BP 166, 38042 Grenoble, France}
\affiliation{$^2$CNRS-Laboratoire de Physique Statistique, Ecole Normale
Sup\'{e}rieure, 24 rue Lhomond, 75231 Paris Cedex 05, France}
\date{\today}

\begin{abstract}
We discuss the quasi two-dimensional trapped Bose gas where the thermal
occupation of excited states in the tightly confined direction is
small but remains finite in the thermodynamic limit.  We show that
the semiclassical theory describes very accurately the density profile
obtained by Quantum Monte Carlo calculations in the normal phase above
the Kosterlitz--Thouless temperature $T_{\KT}$, but differs strongly
from the predictions of  strictly two-dimensional mean-field theory,
even at relatively high temperature.  We discuss the relevance of our
findings for analyzing ultra-cold-atom experiments in
quasi two-dimensional traps.
\end{abstract}

\maketitle
For many years, the physics of two-dimensional quantum gases has been
under close experimental and theoretical scrutiny.  The quest for quantum
phase transitions in two-dimensional atomic Bose gases has started with
experiments on spin-polarized atomic hydrogen adsorbed on liquid $^4$He
surface reaching the quantum degenerate regime \cite{mosk98} and observing
the onset of quasi-condensation \cite{safonov98}.  The Kosterlitz--Thouless
transition \cite{KT} was observed recently in trapped atomic gases of
ultra-cold $^{87}$Rb atoms in an optical lattice potential with a tightly
confined $z$-direction \cite{Dalibard_2006}.

In contrast to experiments with liquid $^4$He films \cite{reppy}, where
the Kosterlitz--Thouless transition is realized directly, the typical
extension in the $z$-direction in the experiments on two-dimensional
gases is much larger than the three-dimensional scattering length.
For this reason, the effective two-dimensional interaction strength
remains sensitive to the density distribution in $z$  \cite{dima2000}.
Nevertheless, the gas is kinematically two-dimensional because of the
strong out-of-plane confinement.

In this letter, we consider the quasi two-dimensional regime
of the trapped Bose gas,  where the temperature $T$ is of
the order of the level spacing in $z$. This corresponds to the
experimental situation with small, but not completely negligible,
thermal occupation of a few excited states of the tightly confining
potential\cite{Dalibard_2006,Dalibard_2007}. The quasi two-dimensional
regime crosses over to the three-dimensional and the strictly
two-dimensional Bose gases as the potential in the $z$-direction is
varied.

It was noticed in experiments \cite{Dalibard_2007,Dalibard_2007b}  and in
direct Quantum Monte Carlo calculations \cite{Holzmann_Krauth_07} that
the density profile of the gas deviates strongly from two-dimensional
mean-field theory, even at relatively high temperature.  We point out
that these deviations can be accounted for by a quasi two-dimensional
mean-field theory which incorporates corrections due to the tightly
confined direction.

We first discuss Bose--Einstein condensation  of the ideal gas in strongly
anisotropic harmonic traps. We define the quasi two-dimensional regime
where the ideal-gas critical temperature is always lower than that of
the strictly two-dimensional ideal gas.  Including interactions on the
level of mean-field, we obtain the density profiles in the semiclassical
approximation and solve the self-consistent mean-field equations
directly. Remarkable agreement of the semiclassical density profiles
with the results of Quantum Monte Carlo calculations is obtained in
the high-temperature normal phase down to the Kosterlitz--Thouless
temperature. The profiles should be very convenient for calibrating
the temperature in experiments of quasi two-dimensional Bose gases.
Comparison of experimental density profiles with Quantum Monte Carlo data
has already removed the original discrepancy of the Kosterlitz--Thouless
temperature between calculation and experiment \cite{Dalibard_2007b}.

We consider an anisotropic trap with oscillator frequencies $\omega \equiv
\omega_x=\omega_y\ll \omega_z$.
At temperature $T\sim  \hbar \omega_z$, the motion is semiclassical
in the coordinates $x$, $y$ and in the momenta $\hbar k_x$, $\hbar
k_y$, whereas the quantization in the $z$-direction is best described
through the energy levels $\nu \hbar \omega_z\ (\nu=0,1,\dots)$ of the
corresponding harmonic oscillator.  Semiclassically, the number $\ddd{N}$
of particles per phase space element $\ddd{k_x}\ddd{k_y}\ddd{x}\ddd{y}$
in the energy level $\nu $ is given by \cite{Landau_Lifshitz}
\begin{equation}
\ddd{N}
= \frac{1}{(2 \pi)^2}
\frac{\ddd{k_x}\ddd{k_y}\ddd{x}\ddd{y}}
{\expc{\beta (\frac{\hbar^2 k^2}{2m} +v(r) + \nu \hbar \omega_z- \mu)}-1},
\elabel{phase_space_density}
\end{equation}
where $\beta=1/T$, $k^2= k_x^2+k_y^2$, and where $v(r)$ is an arbitrary two-dimensional
potential energy (with $r^2=x^2+y^2$).

\eqq{phase_space_density}  can be integrated over all momenta and summed over
all oscillator levels to obtain the two-dimensional particle density
\begin{align}
n(r)  &= \sum_{\nu}\int_0^{\infty} \frac{\ddd{k^2} }{4 \pi}
\frac{1}{\expb{\beta (\frac{\hbar^2 k^2}{2m} +v(r)+ \nu \hbar \omega_z  -\mu)}  -1} \notag\\
         &= - \frac{1}{\lambda^2} \sum_{\nu=0}^{\infty}
           \ln \{1 - \expc{\beta(\mu -v(r)- \nu \hbar \omega_z)}\},
           \elabel{density_r_mu}
\end{align}
where $\lambda = \sqrt{2 \pi \hbar^2 \beta/m}$ is the thermal wavelength.
The potential $v(r)$ can itself contain the interaction with the
density $n(r)$, so that \eq{density_r_mu} is in general a self-consistency
equation.  
The integral of $n(r)$ over space yields the equation of state, that is,
the total number of particles as a function of temperature and chemical
potential.

Let us first consider the ideal gas, where the potential energy $v(r)=m
\omega^2 r^2/2$ is due only to the trapping potential, so that the rhs
of  \eq{density_r_mu} is independent of the density $n(r)$.  We get
\begin{align}
N &= -\frac{\pi}{\lambda^2} \sum_{\nu=0}^{\infty} \int_0^{\infty} \ddd{(r^2)}
 \ln \glc 1 - \exph{\beta(\mu - \nu \hbar \omega_z - m \omega^2 r^2/2)} \grc \notag\\
       &= \frac{T^2}{ \hbar^2\omega^2} \sum_{\nu=0}^{\infty} \F(- \mu \beta + \nu \beta \hbar \omega_z),
\elabel{total_number}
\end{align}
where we have defined
\begin{equation*}
 \F[s](x)= \sum_{n=1}^{\infty} \frac{\exph{-nx}}{n^s }.
\end{equation*}

The saturation number $N_{\SAT}(T)$ is the maximum number of
excited particles (reached at $\mu=0$) at a given temperature.
We have
\begin{equation}
N_{\SAT}^{\text{q2d}} = \frac{T^2}{ \hbar^2 \omega^2} \sum_{\nu=0}^{\infty} \F(\nu \beta \hbar \omega_z).
\elabel{Nsaturation}
\end{equation}
The above relation between the saturation number and the temperature
defines the dependence of the Bose--Einstein condensation temperature
on the particle number $N$.  As mentioned before, the strictly two-dimensional
limit is characterized by the limit $\beta \hbar \omega_z \rightarrow \infty$
(the level spacing in $z$ is much larger than the temperature).
In this limit, only the first term in \eq{total_number} contributes.
Using $\F(0)=\pi^2/6$, we find
\begin{equation}
N_{\SAT}^{\text{2d}}(T) =
\frac{T^2}{\hbar^2 \omega^2} \frac{\pi^2}{6}
\Leftrightarrow \TBECZERO(N) =\frac{\sqrt{6 N}\hbar \omega}{\pi}.
\elabel{critical_temperature_2d}
\end{equation}
In the quasi two-dimensional case, with finite $\beta \hbar \omega_z $,
the occupation of the oscillator levels $\nu=1,2,\dots$ increases
the saturation number and therefore lowers the critical temperature.
It is convenient to express in units of $\TBECZERO$ both the
temperature $t= T/\TBECZERO$ and the oscillator strength
$\omegatilde_z= \hbar \omega_z/\TBECZERO$, and to write $\mutilde = \beta \mu$.
Using \eq{critical_temperature_2d}, we may rewrite the equation of
state, \eq{total_number},  as a relation between the temperature
$t$, the chemical potential $\mutilde$, and the oscillator strength
$\omegatilde_z$,
\begin{gather}
t = f(t,\mutilde,\omegatilde_z), \elabel{self_consistency_temp}\\
\intertext{with}
f(t,\mutilde,\omegatilde_z) = \left(\frac{6}{\pi^2}\sum_{\nu=0}^{\infty}
\F  \glb -\mutilde + \nu \omegatilde_z/t\grb \right)^{-1/2}.
\elabel{iteration_ideal}
\end{gather}
\eqq{self_consistency_temp} is solved numerically by iterating
$t_{n+1}=f(t_n)$ from an arbitrary starting temperature $t_0$
to the fixed point. The critical temperature $t_\BEC = \TBECQZERO/\TBECZERO$ (as
a function of $\omegatilde_z$) of the quasi two-dimensional ideal
Bose gas is the solution for $\mutilde=0$ (see \fig{BEC_ideal_quasi}).  The reduction
with respect to the strictly two-dimensional case
is notable for systems of  experimental interest. For example, we
find $t_\BEC=0.78$ for the experimental value $\omegatilde_z=0.55$
\cite{Dalibard_2006, Dalibard_2007} considered in
the Quantum Monte Carlo calculations \cite{Holzmann_Krauth_07}.
\begin{figure}
   \epsfig{figure=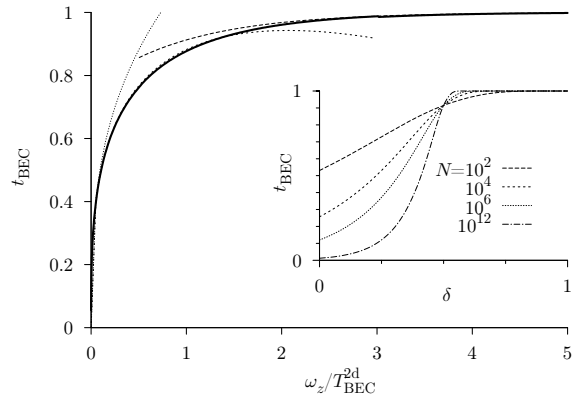,width=0.5\textwidth}
   \caption{Bose--Einstein condensation temperature $t_\BEC$  of the ideal
   quasi two-dimensional gas in a harmonic trap with $ \omega_z/\omega
   \propto N^{1/2}$ (expansions from \eq{expansion_of_t}).   The inset
   shows $t_\BEC$ for scaling $\omega_z/\omega = N^{\delta}$ as a
   function of $\delta$ for different $N$. At $\delta=0$, the critical
   temperature $\TBECZEROTHREE= [N/\zeta(3)]^{1/3}$ of the ideal gas in
   a three-dimensional isotropic trap is recovered.}
\flabel{BEC_ideal_quasi}
\end{figure}

We can expand \eq{iteration_ideal} for small and for large $\omegatilde_z$
and find
\begin{align}
t_\BEC \sim
\begin{cases}
 \glc \frac{\zeta(2)}{\zeta(3)}\grc^{1/3}\omegatilde_z^{1/3}-\frac{1}{6}\frac{\zeta(2)}{\zeta(3)}
\omegatilde_z &\text{for $\omegatilde_z \ll 1$}\\
1 - \frac{1}{ 2 \zeta(2)^{3/2}} \expb{-\omegatilde_z}   &\text{for $\omegatilde_z \gg 1$}
\end{cases}, 
\elabel{expansion_of_t}
\end{align}
where we have used $\zeta(s) \equiv F_s(0)$ (note that
$\zeta(2)=\pi^2/6$ and $\zeta(3) \simeq 1.202$).  The expansions
are indicated in  \fig{BEC_ideal_quasi}. They give the critical
temperature to better than $1.2 \%$ for all values of $\omegatilde_z$
(the low-$\omegatilde_z$ expansion is used for $\omegatilde< 1.8$ and
the high-$\omegatilde_z$ expansion for $\omegatilde> 1.8$).  The first
term in the small-$\omegatilde_z$ expansion of \eq{expansion_of_t}
corresponds to the three-dimensional gas.  Indeed, $t_{\BEC}\sim
[\zeta(2)/\zeta(3)]^{1/3} \omegatilde_z^{1/3}$ is equivalent to
\begin{align}
   T_{\BEC} &\sim \glc \frac{\zeta(2)}{\zeta(3)} \grc^{1/3} (\hbar \omega_z)^{1/3} 
                        (\TBECZERO)^{2/3} \notag\\
            & \sim \glc \hbar ^3 \omega_z \omega^2/\zeta(3) \grc^{1/3} N^{1/3},
   \elabel{two_d_three_d}
\end{align}
the well-known condensation temperature for the three-dimensional
Bose--Einstein gas in an anisotropic trap \cite{Kleppner}.
\eqq{two_d_three_d}  also follows directly from \eq{Nsaturation}
by replacing the sum over the oscillator levels by an integral.
Likewise, the first term in the large-$\omegatilde_z$ expansion of
\eq{expansion_of_t} represents the strictly two-dimensional gas.

The inset of \fig{BEC_ideal_quasi} further analyzes the expansions of
\eq{expansion_of_t}. Indeed, we can choose a scaling $\omega_z/\omega \sim
N^{\delta}$ different from the quasi two-dimensional case $\delta=1/2$.
Any choice of $\delta  < 1/2$ corresponds to $\omegatilde_z \to 0$
for $N \to \infty$ so that asymptotically the three-dimensional
regime is reached.  Analogously, $\delta > 1/2$ corresponds to
$\omegatilde_z \to \infty$ for $N \to \infty$, driving the transition
into the strictly two-dimensional regime. The rescaled
transition temperatures are plotted for $\omega_z/\omega = N^{\delta}$ where the
case $\delta=0$ corresponds to the three-dimensional isotropic trap
\cite{footnote_direct_evaluation}.

In order to  describe interaction effects in the quasi two-dimensional
Bose gas, we now add a semiclassical contact term to the potential energy
of \eq{density_r_mu}:
\begin{equation}
v(r)=m \omega^2 r^2/2 + 2 g [n(r) - n(0)].
\elabel{effective_potential}
\end{equation}
We have subtracted the central density, so that the potential still
vanishes at the origin.  As discussed earlier \cite{Holzmann_Krauth_07},
the effective interaction $g= 4 \pi a \hbar^2/m \int \dd{z} [\rho(z)]^2$
is proportional to the three-dimensional s-wave scattering length $a$
and to the integral of the squared density distribution in $z$, described
by the normalized diagonal density matrix $\rho(z)$.  In the temperature
range of interest, this density distribution is well described by the
single-particle harmonic-oscillator density matrix in $z$, leading to
\begin{equation}
   g= a\sqrt{\frac{8 \pi \omega_z \hbar^3}{m}}  
       \sqrt{\tanh[\omegatilde_z/(2t)]}.
   \elabel{g_effective}
\end{equation}
The effective interaction thus decreases with temperature from its
zero-temperature value $\gtilde=a \sqrt{8 \pi \omega_z\hbar^3/ m}$.
To keep the interaction strength fixed, we must keep $g$ (or equivalently
$\gtilde$) constant in the quasi two-dimensional thermodynamic limit
which requires a fixed value of the scaled scattering length 
$a \sqrt{\omega_z}\propto a \omega^{1/2} N^{1/4}$.

The mean-field density $n(r)$, on the lhs of \eq{density_r_mu}, depends on the
variable $r$ only via the potential $v(r)$.  In the space integral over
the particle density, we can thus change the integration variable from $r$ to $v$. This
allows us to determine the equation of state explicitly:
\begin{multline}
   N = \pi \int_0^{\infty} \dd{(r^2)} n(r) 
   =  \pi \int_{0}^{\infty} \dd{v} \glc \frac{\partial (r^2)}{\partial v}
   \grc n(v) \\ = \frac{2\pi}{m \omega^2}
    \int_{0}^{\infty} \dd{v} \glc 1 - 2 g \frac{\partial n}{\partial v}
    \grc n(v)\\
   =\! \frac{T^2}{\hbar^2 \omega^2}\!\! \glb
    \sum_{\nu=0}^{\infty} \F(-\mutilde + \nu\! \beta \hbar \omega_z)
   +\! \frac{ m g}{2 \pi \hbar^2}\! \glc n(0) \lambda^2 \grc^2 \grb\! ,
   \elabel{V_eff_transformation} 
\end{multline} 
where the central density,
\begin{equation}
   n(0)\lambda^2= - \sum_{\nu=0}^{\infty} \ln \gld 1 - \expb{
   \mutilde - \nu \omegatilde_z/t} \grd, 
   \elabel{n0mutilde}
\end{equation} 
is directly expressed in terms of $\mutilde$, independent of the interaction, due
to the subtraction performed in our effective potential,
\eq{effective_potential}. (Note that the first integral on the second line
of \eq{V_eff_transformation} has already appeared in \eq{total_number}
and that the second integral is a total derivative.)
The equation of state can be written in terms of the temperature $t$.
This yields the following  generalization of \eq{iteration_ideal} to the 
mean-field gas:
\begin{gather}
t = \! \!\glb\!\! \frac{6}{\pi^2}\!\! \left[\sum_{\nu=0}^{\infty} \F(- \mutilde 
\! +\! \nu \frac{\omegatilde_z}{t})
+\! \frac{mg}{2 \pi \hbar^2} [n(0) \lambda^2]^2 \right]\! \grb^{-1/2} \!\!.
\elabel{iteration_mean_field}
\end{gather}
An iteration procedure $t_{n+1} = f(t_{n})$ again 
obtains the temperature $t = T/\TBECZERO$ as a function of
the chemical potential $\mutilde$ for given parameters $\omegatilde_z$
and $g$. (The central density is computed using \eq{n0mutilde}
during each iteration.)

The mean-field density profile is obtained in two steps by first
calculating the density profile as a function of the scaled effective
potential $\vtilde= \beta v$,   
\begin{equation*}
n(\vtilde) \lambda^2= - \sum_{\nu=0}^{\infty} \ln 
   \gld 1 - \expb{\mutilde- \vtilde - \nu \omegatilde_z/t} \grd, 
\end{equation*}
and then by inverting
\eq{effective_potential} to obtain $r(n)$ (thus $n(r)$) for the given  
$\vtilde$ and $n\lambda^2$:
\begin{equation}
r(n\lambda^2, \vtilde)= \sqrt{ \frac{2 T }{m \omega^2} \glb 
 \vtilde - \frac{mg n\lambda^2}{\pi\hbar^2} \grb}.
\elabel{Vtilde_def}
\end{equation}
\begin{figure}
\epsfig{figure=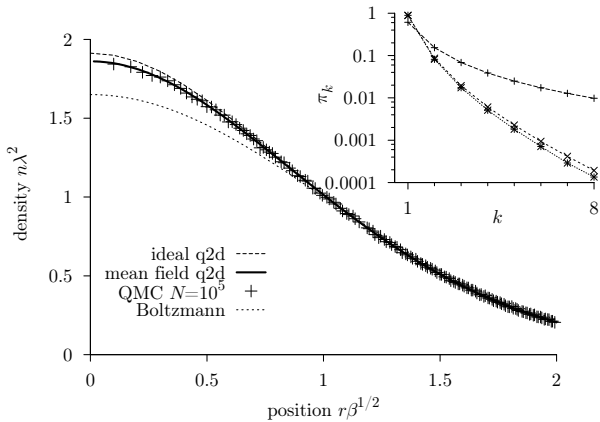,width=0.5\textwidth}
   \caption{Two-dimensional density profile $n(r= \sqrt{x^2 + y^2})
   \lambda^2$ at temperature $T = \TBECZERO$ in a trap with $\omega_z=0.55
   \TBECZERO$, in the ideal Bose gas and for $m\gtilde/\hbar^2=0.13$ according to mean-field
   theory, compared to Quantum Monte Carlo simulations at
   $N=10\,000$, and for ideal distinguishable particles (first term
   in \eq{gaussian_fit}). The inset shows the cycle weights $\pi_k$ for
   the strictly two-dimensional and the quasi two-dimensional Bose gas,
   and for the Quantum Monte Carlo simulations (from above).}
   \flabel{profile_1}
\end{figure}
In \fig{profile_1}, we show the remarkable agreement of the quasi
two-dimensional mean-field profile with the one obtained by Quantum Monte
Carlo simulations as in Ref.~\cite{Holzmann_Krauth_07}, for $N=10\,
000$ bosons for parameters $\omegatilde_z=0.55$ $t=T/\TBECZERO=1$,
and $m\gtilde/\hbar^2=0.13$.  The simulations take into account
the full three-dimensional geometry, and particles interact via the
three-dimensional s-wave scattering length $a$ (see Ref.~\cite{qmc}
for a more detailed description of finite temperature simulations of
trapped Bose gases).  Comparison with the ideal quasi two-dimensional
gas is also very favorable.

The mean-field density is an exact sum of Gaussians,
\begin{equation}
n(r)\lambda^2 =\frac{\pi^2}{6t^2} \sum_{k=1}^{N} k \pi_k  
\expc{- \frac{m \omega^2 (r\sqrt{k\beta})^2}{2}}, 
\elabel{gaussian_fit}
\end{equation}
whose variances correspond to the density distribution of the harmonic
oscillator at temperature $k \beta$.  The pre\-factors  in \eq{gaussian_fit}
contain the cycle weights $\pi_k$.  These weights give the probability
of a particle to be in a cycle of length $k$ in the path-integral
representation of the Bose gas, where the density matrix must be
symmetrized through a sum of permutations\cite{Chevallier,SMAC}.
For the ideal Bose gas, the $\pi_k$ are easily computed. In the
distinguishable-particle limit, at infinite temperature, only cycles of
length $k=1$ contribute ($\pi_1=1$), whereas the Bose--Einstein condensate
is characterized through contributions of cycles of length $k \propto N$.
The cycle weights $\pi_k$ are shown in the inset of \fig{profile_1}
for small $k$.  Only very short cycles contribute, and the density
profile can thus be described by a very small number of Gaussians.
The exact cycle-weight distribution of the Quantum Monte Carlo does
not rigorously correspond to a profile as in \eq{gaussian_fit}, although
the corrections are negligible in our case.

\begin{figure}
\epsfig{figure=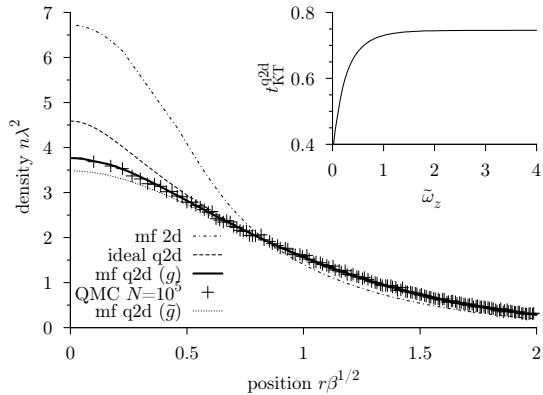,width=0.5\textwidth}
   \caption{Two-dimensional density profile $n(r= \sqrt{x^2 + y^2})
   \lambda^2$ at temperature $T = 0.8\,\TBECZERO$ in a trap with
   $\omega_z=0.55 \TBECZERO$, in the ideal Bose gas  and for
   $m\gtilde/\hbar^2=0.13$ according to strictly two-dimensional and 
   quasi two-dimensional mean-field theory (using $g$ and $\gtilde$), 
   compared to Quantum Monte Carlo simulations with $N=10\,000$.
   The inset shows the Kosterlitz--Thouless temperature within
   modified mean-field theory as a function of $\omegatilde_z$ (
   the strictly two-dimensional limit is  $t_{\KT}^{\text{2d}}=0745$ 
   (from \eq{t_crit_2d}).}
   \flabel{profile_0.8}
\end{figure}

\figg{profile_0.8} considers the temperature $t=0.8$  in the
interval between the Kosterlitz--Thouless temperature of the quasi
two-dimensional interacting gas (at $t=t_{\KT}\simeq 0.70$ for these
parameters \cite{Holzmann_Krauth_07}) and the strictly two-dimensional
Bose--Einstein condensation temperature.  Again, the  agreement of
the quasi two-dimensional mean field with the  exact density profile
obtained by Quantum Monte Carlo is remarkable. At this temperature,
the deviations with the ideal quasi two-dimensional profile and with
the strictly two-dimensional mean field are important.  To illustrate
the temperature-dependence of the effective interaction, we also show
the mean-field profile computed with the zero-temperature interaction
parameter $\gtilde$ instead of the true effective two-dimensional
interaction $g$ (see  \eq{g_effective}).  We note in this context
that the difference in \eq{g_effective} between $g$ and $\gtilde$ was
determined under the condition that the interaction leaves the density
distribution in $z$ unchanged. Whenever this condition is violated,
the density distribution in $z$ must be computed by other means, as for
example by Quantum Monte Carlo methods (see \cite{Holzmann_Krauth_07}).

Let us finally discuss the Kosterlitz--Thouless transition into the
low-temperature phase, which is not contained in mean-field theory.
The semiclassical quasi two-dimensional gas does not Bose-condense
because the particle number in \eq{V_eff_transformation} diverges  at
$\mutilde=0$ (it saturates at a finite value in the ideal Bose gas).
This divergence is due to the logarithmic divergence of
the central density (see \eq{n0mutilde}).  However, interaction effects
beyond mean-field drive a Kosterlitz--Thouless phase transition \cite{KT}
from the high-temperature normal phase to a superfluid below $T_{\KT}$.

As discussed previously \cite{TKTB,Holzmann_Krauth_07}, the
Kosterlitz--Thouless transition occurs when the central density $n(0)
\lambda^2$ reaches the critical value of the two-dimensional homogeneous
gas, which has been determined numerically \cite{Svistunov2D} for $g \to 0$:
\begin{equation}
   n(0) \lambda^2 \simeq n_c \lambda^2 \simeq \log \frac{380 \hbar^2}{mg}.
   \elabel{critical_density}
\end{equation}
We can introduce (by hand) the concept of a critical density  into
mean-field theory by selecting among the solutions $t(\mutilde)$ of
\eq{iteration_mean_field} the one satisfying \eq{critical_density}. For
the interaction parameters used in Ref \cite{Holzmann_Krauth_07}, $m
\gtilde/\hbar^2=0.13$, we find a mean-field critical temperature $t_\KT^{\text{q2d}}
=T_{\KT}^{\text{q2d}}/\TBECZERO=0.69$.
This value is  in excellent agreement with the Monte Carlo data. The
inset of \fig{profile_0.8} shows the variation of this 
mean-field critical 
temperature as a function of $\omegatilde_z$ (for $m\tilde{g}/\hbar^2=0.13$).

The calculation of the mean-field critical temperature simplifies further
in the strictly two-dimensional Bose gas, because the chemical potential
in \eq{n0mutilde} is then an explicit function of the critical density,
and can be entered into \eq{V_eff_transformation}.  With $\mutilde_c=
\ln \gld 1 - \expb{-n_c \lambda^2} \grd \simeq - mg/(380 \hbar^2)$, and
by again transforming the equation for $N$ vs. density into a relation
between critical temperatures \cite{TKTB,Holzmann_Krauth_07}, we obtain
\begin{equation}
t_\KT^{\text{2d}} = 
\frac{T_{\KT}^{\text{2d}}}{\TBECZERO}  = \glc 1 + \frac{3 m g}{\pi^3 \hbar^2} \glb  \ln
\frac{380 \hbar^2}{mg}\grb^2 \grc^{-1/2},
\elabel{t_crit_2d}
\end{equation}
where we have neglected small corrections of order $\mutilde_c
\log |\mutilde_c|$.  The strictly two-dimensional limit
$t_\KT^{\text{2d}}=0.745$ for $mg/\hbar^2=0.13$ agrees with the data shown
in the inset of \fig{profile_0.8} in the large $\omegatilde_z$ limit.

In conclusion, we have considered in this letter the semiclassical
description of the quasi two-dimensional trapped Bose gas.  We have
compared this description with Quantum Monte Carlo data and  have shown
that the density profiles are accurately reproduced in the normal phase
down to the Kosterlitz--Thouless temperature. The thermal occupation of
excited states in the out-of-plane direction quantitatively explains the
large deviations of the density profiles from strictly two-dimensional
mean-field theory, which was recently noticed in the experiment
\cite{Dalibard_2007}. Originally, it was speculated that the emerging
almost-Gaussian density profiles could be attributed to effects beyond
mean-field \cite{Dalibard_2007,Dalibard_2007b}. However, even though
the thermal occupation of the excited states in the tightly confined
direction is clearly noticeable  for  the experimental parameters,
the transition itself is still of the Kosterlitz--Thouless type as
revealed by the experimental coherence patterns \cite{Dalibard_2006}
and confirmed by numerical calculations of the algebraic decay of the
condensate density with system size \cite{Holzmann_Krauth_07}.

\section{acknowledgments}
We are indebted to Jean Dalibard for many inspiring discussions.


\begin{thebibliography}{99}
\bibitem{footnote_direct_evaluation} The inset of \fig{BEC_ideal_quasi} is
evaluated from the asymptotic expansion \eq{expansion_of_t}. Very similar
results for $T_c$ follow from the exact saturation numbers at finite $N$.
\end{thebibliography}
\end{document}